\documentclass[9pt,twocolumn,twoside]{pnas-new}
\usepackage{mathtools}

\templatetype{pnasresearcharticle} 

\title{Optimizing Brownian escape rates by potential shaping}

\author[a,b]{Marie Chupeau\textsuperscript{1}}
\author[c]{Jannes Gladrow\textsuperscript{1}}
\author[d]{Alexei Chepelianskii}
\author[c]{Ulrich F. Keyser}
\author[a]{Emmanuel Trizac\textsuperscript{2}}

\affil[a]{LPTMS, CNRS, Univ. Paris-Sud, Universit\'e Paris-Saclay, 91405 Orsay, France }
\affil[b]{Magic LEMP, Centrale-Supélec, 91190 Gif, France}
\affil[c]{Cavendish Laboratory, University of Cambridge, Cambridge CB3 0HE, United Kingdom}
\affil[d]{LPS, CNRS, Univ. Paris-Sud, Universit\'e Paris-Saclay, 91405 Orsay, France}

\leadauthor{Chupeau} 

\significancestatement{Diffusion is of fundamental relevance in physics, chemistry and biology,
to explain the transport of macromolecules in a solvent, under
the action of external forces. Diffusion is fueled by temperature,
resulting in thermal fluctuations. Macromolecules
trapped in a potential energy well, can then escape,
harnessing favorable fluctuations. Such events are a priori rare
nevertheless, and the so-called Arrhenius relation describes the
corresponding exponential decrease of escape rates with increasing barrier
heights. It therefore comes as a surprise that some potential wells, with
high barriers, may help the macromolecule escape. We report the first
experimental proof of this counterintuitive boosting effect. Optimizing theoretically the potential profile that provides maximum
boost, we translate our predictions into results in automated
microfluidic experiments.}

\authorcontributions{ MC, JG, AC, UK and ET designed research and wrote the paper. JC and UK performed the experimental work while MC, AC and ET performed the theoretical analysis.}
\authordeclaration{No conflict of interest.}
\equalauthors{\textsuperscript{1}M.C.(Author One) and J.G. (Author Two) contributed equally to this work.}
\correspondingauthor{\textsuperscript{2}To whom correspondence should be addressed. E-mail: trizac@lptms.u-psud.fr }


\begin{abstract}
Brownian escape is key to a wealth of physico-chemical processes, including polymer folding, and information storage. 
The frequency of thermally activated energy barrier crossings is assumed to generally decrease exponentially with increasing barrier height. Here, we show experimentally that higher, fine-tuned barrier profiles result in significantly enhanced escape rates in breach of the intuition relying on the above scaling law, and address in theory the corresponding conditions for maximum speed-up. Importantly, our barriers end on the same energy on which they start. For overdamped dynamics, the achievable boost of escape rates is, in principle, unbounded so that the barrier optimization has to be regularized. We derive optimal profiles under two different regularizations, and uncover the efficiency of  N-shaped barriers. We then demonstrate the viability of such a potential in automated microfluidic Brownian dynamics experiments using holographic optical tweezers and achieve a doubling of escape rates compared to unhindered Brownian motion. Finally, we show that this escape rate boost extends into the low-friction inertial regime.
\end{abstract}

\dates{This manuscript was compiled on \today}
\doi{\url{www.pnas.org/cgi/doi/10.1073/pnas.XXXXXXXXXX}}

\begin{document}

\maketitle
\thispagestyle{firststyle}
\ifthenelse{\boolean{shortarticle}}{\ifthenelse{\boolean{singlecolumn}}{\abscontentformatted}{\abscontent}}{}

\dropcap{A}rrhenius law, a key principle of reaction kinetics, posits that chemical reactions become exponentially slower, the higher the activation energy barrier that reactants have to overcome\footnote{Arrhenius himself attributed this law to van't Hoff~\cite{Arrhenius,Hanggi}.}.
In 1940, Kramers published a comprehensive theory for Arrhenius' scaling, introducing a framework for thermally activated transitions in an energy landscape. Importantly, in his theory, the system is coupled to the environment through friction and thermal noise. Further research has since revealed that swift thermal escapes from local potential energy minima require an intermediate friction magnitude such that motion is neither sluggish nor deterministic~\cite{Kramers, Hanggi,Melnikov,Lutz,Rondin}. 
However, influences of barrier shapes on escape rates and conditions of optimality thereof have been hitherto overlooked in the literature.
In this letter, we theoretically optimize static barrier profiles, calculate the corresponding speed-limit of escape, and demonstrate experimentally that higher, optimized barriers paradoxically result in increased escape rates, in contrast to intuition based on Kramers law. 
Since the maximum achievable escape rate is infinite, the barrier optimization has to be constricted, e.g. by placing an upper bound on the barrier height or curvature. 
In addition, we demonstrate experimentally a doubling of escape rates compared to unhindered Brownian motion, which proves that our predicted barrier profiles can indeed be realized. 
Furthermore, we show that the rate-boost applies over a range of friction values, extending from the overdamped into the inertial regime. 
Our results indicate that fine-tuned free-energy landscapes of higher amplitude may increase reaction rates. In the context of protein folding, a carefully rate-optimized free-energy landscape may thus well exhibit a larger number of intermediate states in spite of additional necessary escapes~\cite{Wagner:1999}.  

We believe that this paper will invigorate a search for Brownian optimality, and inform the design of systems where thermal excitation plays a role, such as adatom diffusion \cite{Guantes}, chemical dynamics \cite{GarciaMuller}, polymer folding \cite{Best}, and magnetic information storage where thermal fluctuations limit capacity~\cite{Coffey, Rotskoff2017}. The question of optimizing the 
potential profile becomes timely in view of the spectacular experimental progress made in
controlling confining features for colloidal objects \cite{Padgett,martinez_engineered_2016,ciliberto_experiments_2017}. 

The rate of progress of Brownian or other stochastic processes is not easily quantifiable. One way to measure the ''speed'' of Brownian motion is the mean first-passage time (FPT) to a given distance \cite{K05,Redner}. In Kramers' escape problem, the reciprocal of the escape rate corresponds to the time of first-passage to leave the initial state. A lower bound for the achievable FPT, e.g. of the reaction coordinate of a folding molecule, therefore corresponds to a speed-limit of the ensemble reaction rate~\cite{Kubelka2004}. 

\begin{figure}[h]
\centering
\includegraphics[width=87mm]{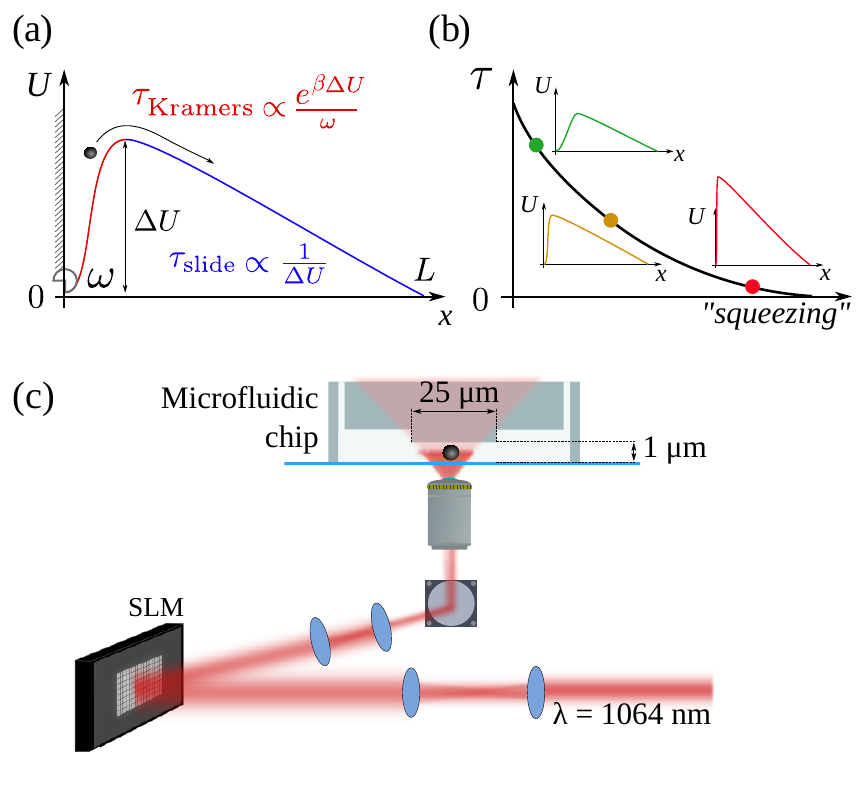}
\caption{(color online) (a) Simple potential profile, characterized by its height $\Delta U$ and the curvature of the associated potential well $\omega$. The two key timescales in the overdamped exit process are the Kramers' escape time$^{\P}$
and the slide time. For such a profile and provided that $\Delta U$ is large enough,
the exit time simplifies into $\tau\simeq\tau_{\rm Kramers} + \tau_{\rm slide}$.
Increasing $\Delta U$ (to decrease $\tau_{\rm slide}$) and $\omega$ in the appropriate fashion leads to enhanced ``squeezing'' of the well, 
and to a decrease of $\tau$, see panel (b). In doing so, $\Delta U$ diverges while $\tau$ can be made as small as desired.
Thus, imposing the constraint $|U(x)|<U_{\rm max}$ or discretizing space will lead to a non-zero optimal time $\tau$
(cases A and B below). This regularizes
the vanishing of $\tau$, that is a specificity of the overdamped description. The underdamped regime does not require regularization\protect\footnote{We generally understand ``underdamped'' as meaning ``non-overdamped''. The overdamped regime is such that there are infinitely many potential shapes that lead to the optimal escape time $\tau=0$. A constrained (regularized) problem is such that $\tau \neq 0$.}. 
(c) Sketch of the holographic optical tweezers setup used to measure escape times over optimized barriers that are shaped by creating intensity and phase profiles inspired by predicted optimal potential profiles. }
\label{fig1}
\end{figure}

\section*{Introductory Example}
In order to illustrate the speed-up of the mean FPT across potential barriers, we consider the triangular barrier profile in Fig.~\ref{fig1}(a). This profile has been shown to decrease first-passage times of overdamped Brownian motion, relative to a linear potential~\cite{Palyulin:2012}. Importantly, this {\bf profile-induced speed-up} does not require any expense in energy\footnote{If ``position'' $x$ refers to a reaction coordinate subsuming a complex landscape, the free energy, rather than the energy, should be considered. This distinction is immaterial for our one-dimensional discussion.}
: the effect also appears in energy-neutral potential profiles, constructed in such a way that initial and final energy levels coincide\footnote{In non-neutral (tilted) landscapes  where net work is performed, interesting effects have been reported.  For instance, diffusion coefficients in tilted periodic potentials can vastly exceed their free diffusion value, which leads to a ``{\em giant acceleration}'' of diffusion~\cite{Reimann2001, Lee2006, Kim2017}.}.
The particle in Fig.~\ref{fig1} (a) is initialized at $x=0$ in the narrow well (red region), and our interest goes to the first-passage at $x=L$. The movements are bounded by a reflecting barrier at $x=0$ and an absorbing boundary at $x=L$. The mean escape time from the narrow well is given by Kramers' result $\tau_{\text{Kramers}}\propto e^{\beta \Delta U}/\omega$
with $\beta=1/(k T)$ denoting the inverse temperature times the Boltzmann constant $k$\footnote{Assuming the bottom of the well together with the top of the barrier to be parabolic, with respective curvatures $m\omega^2$ and $-m\omega^2_{\text{top}}$ where $m$ is the particle mass, Kramer's escape time reads \cite{BarratHansen} $\tau_{\text{Kramers}} \propto \omega^{-1} \omega_{\text{top}}^{-1} \exp[\Delta U /(kT)]$, $\Delta U >0$.}.
Once out of the well, the particle slides towards the exit within an average slide time $\tau_{\text{slide}}\propto 1/\Delta U$ as follows from gradient descent. 
In the limit of high barriers, the overall mean exit time $\tau$ reads as the sum of $\tau_{\text{Kramers}}$ and $\tau_{\text{slide}}$~\cite{Palyulin:2012};
it can therefore be made arbitrarily small by simultaneously increasing the steepness and height of the initial well, see Fig.~\ref{fig1}(b). 
Crucially, a sufficiently high and steep barrier yields a mean exit time shorter than the corresponding free diffusion time $\tau_\text{free} = L^2/(2D)$, where $D$ is the diffusion coefficient~\cite{Gardiner2009}. Moreover, there is no lower bound (other than 0) for the exit time: further ``{\em squeezing}'' will further decrease $\tau$ (see Fig.~\ref{fig1}(b)). The exit time approaches zero for appropriately chosen diverging curvature and barrier height. Fig.~\ref{fig1}(c) 
present the experimental setup used in this work to test the predictions.

\section*{Barrier profile optimization}

 In the following, all relevant quantities, the mean exit time, the potential, and the abscissa are conveniently rescaled: $\widetilde{\tau}=D\tau/L^2$, $\widetilde{U}= \beta U$, $\widetilde{x}=x/L$, where $D=1/(\beta m\gamma)$, defines the  temperature $T$ that drives the Brownian process, $m$ the particle mass and $\gamma$ the friction coefficient. Tildes denote dimensionless variables, but will be dropped hereafter, unless otherwise stated.
The mean exit time is given by~\cite{VanKampen,BarratHansen}
\begin{equation}\label{tau1Dresc}
\tau = \int_0^1 dx \; e^{-U(x)} \int_x^1 dy \; e^{U(y)},
\end{equation}
which is invariant under the transformation $U(x)$ to $-U(1-x)$ (see suppl.~\cite{SM}).
A joint use of this invariance and Cauchy-Schwarz inequality shows that the optimal potential is necessarily antisymmetric with respect to $x=1/2$ (such that $U(x)=-U(1-x)$), provided that the constraints on the potential are compatible with antisymmetry transformation (see suppl.~\cite{SM} for details). The two distinct constraints we discuss in the following, (A) bounds on $U$ or (B) regular space-discretization, are both compatible with antisymmetry.
\subsection*{Constraint A - Symmetrically bound potential}
Imposing bounds on the potential offers the most straightforward regularization to discuss optimality. For the sake of simplicity, we restrict ourselves to constant bounds $U_{\rm min} \leqslant U(x) \leqslant U_{\rm max}$, and refer to this constraint as A.

\begin{figure}[h]
\centering
\includegraphics[width=87mm]{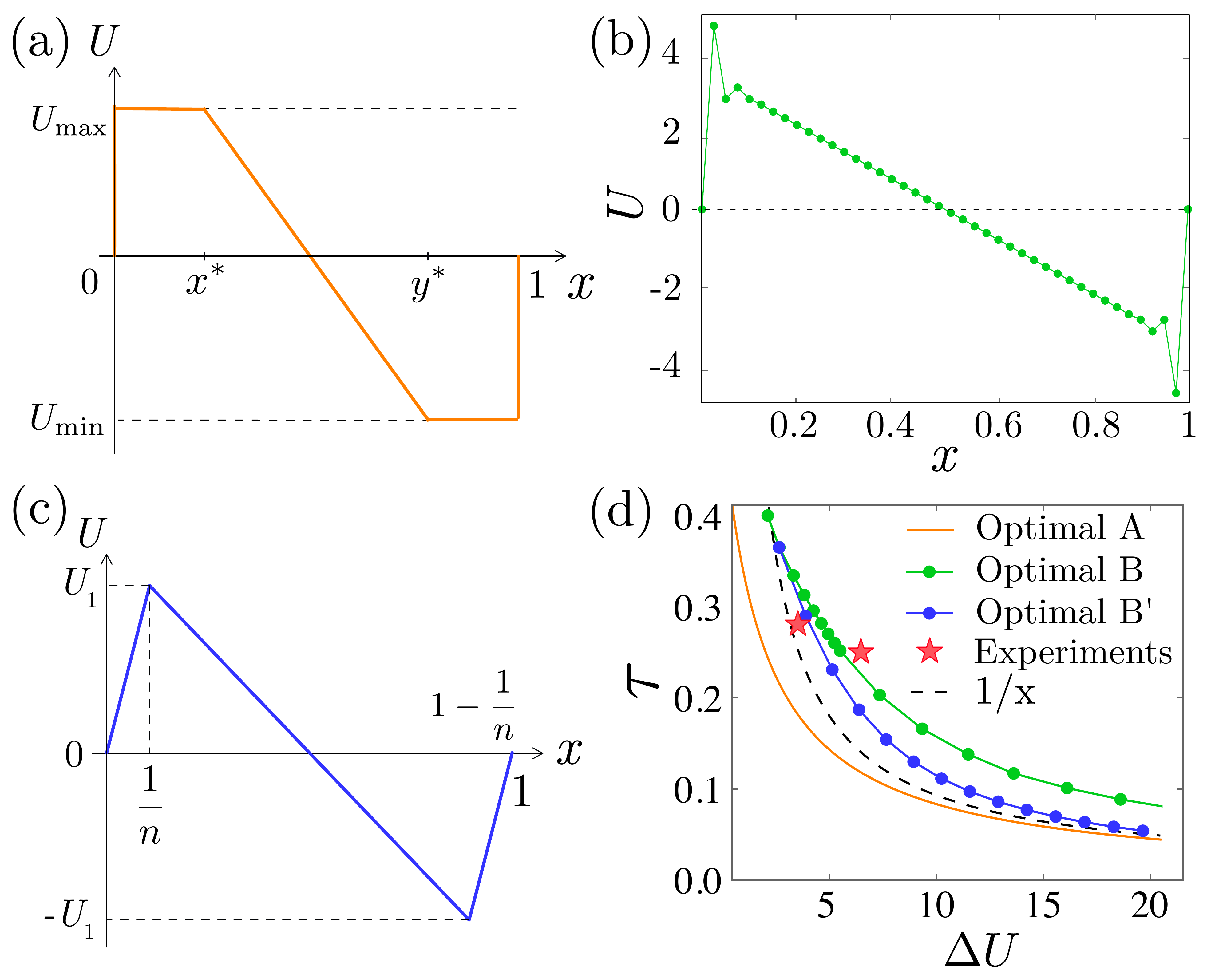}
\caption{(color online) (a) Optimal potential profile A with bounds $U_{\rm min}$ and $U_{\rm max}$. (b) Optimal potential B for a 40-support potential. The potential is not bounded anymore. (c) N-shaped approximation of the discretized case, with only one variable parameter $U_1$ (optimal potential B'). (d) Mean exit time as a function of the potential barrier, for the optimal potentials A (a), B (b) and B' (c). The color code is consistent between the four panels. The dashed line represents the asymptotics $1/\Delta U$, valid for optimal potentials A and B'. Because of its overshoots, the optimal potential B has a different asymptotics. The red stars correspond to the experimental data (see Fig.~\ref{fig3}).}
\label{fig2}
\end{figure}

By minimizing $\tau$ in~\eqref{tau1Dresc} with respect to $U$, 
we show in the supplement~\cite{SM} that the optimal potential A obeys
\begin{equation}\label{var1}
e^{U(y)} \int_0^y dx \, e^{-U(x)} = e^{-U(y)} \int_y^1 dx \, e^{U(x)}.
\end{equation}
It follows that it has the generic shape sketched in Fig.~\ref{fig2}~(a), which consists of two plateaus on the upper and lower bounds, connected by a decreasing linear part. The position 
$x^*$ and $y^*$ of the intersection between the two plateaus and the linear part can also be calculated, as well as the associated optimal mean exit time
\begin{equation}\label{optshape}
\tau_{\rm opt}^A = x^* = 1 - y^* = \frac{1}{2 + U_{\rm max} - U_{\rm min}}.
\end{equation}
For symmetric bounds $U_{\rm min}=-U_{\rm max}$, the optimal potential is antisymmetric, as expected. For general bounds, the optimal mean exit time only depends on the potential difference $\Delta U = U_{\rm max}-U_{\rm min}$ and is always smaller than the free diffusion time $\tau_{\rm free}=1/2$. Moreover, when the potential difference is much larger than one, we obtain
\begin{equation}\label{Heisenberg}
\tau^A_{\rm opt} \Delta U \simeq 1.
\end{equation}
This expression is reminiscent of Heisenberg's time-energy uncertainty principle, even though the problem is, of course, purely classical. The larger the amplitude of the potential, the shorter the mean exit time. \eqref{Heisenberg} also implies that the scaling of the mean exit time reduces, at leading order, to the slide time-scaling $1/\Delta U$\footnote{Going back to dimensioned quantity, this corresponds to a speed $\tau/L \propto U/(m\gamma L)$, where temperature drops out. This is nothing but the sliding time in the constant force field $U/L$, with a mobility $1/(m \gamma)$. The initial escape from $x=0$ to the plateau at $x=0^+$, (and likewise the jump from $x=L^-$ to $x=L$) occur in a vanishing time, within the present overdamped formulation.}.
This result is in accord with the fact that the plateaus disappear at large $\Delta U$. This phenomenology also applies in higher dimensional systems, as shown in the supplement~\cite{SM}.

Despite its convenience, constraint A possesses one drawback: the non-physical discontinuity of the corresponding optimal potential A\footnote{$U(0)=U(1)=0$ by construction, while $U(0^+)= U_{\rm max}$ and $U(1^-)=U_{\rm min}$}.

\subsection*{Constraint B - Piecewise-linear potential}

We therefore turn to constraint B, for which the potential is a piecewise linear function, defined by $n$ nodes ($x_i,U_i$) where we require the $x_i$ to be regularly spaced in $[0,1]$. We refer to such a profile as a $n$-support function. As before, the potential is chosen to be energy-neutral, that is, $U_0=U_{n+1}=0$. 
Contrarily to constraint A, constraint B does not impose any restriction on the value of $U_i$ ($U$ is {\it a priori} not bounded). 
As shown in the introductory example, large potential barriers can only be efficient when the width of the associated well vanishes. Here, the well width is bounded from below by $1/n$, the spatial discretization step, so that bounding the potential becomes unnecessary.  
In order to compute the associated optimal potential profile B, we carry out simulated annealing. Fig.~\ref{fig2}~(b) provides an example of optimal potential B for a 40-support. It is antisymmetric as expected, and reminiscent of an ``N-shape'' with an overshoot and an undershoot on both sides of the intermediary slide. The overshoot prevents the particle from recrossing the barrier and falling back into the initial well. Its amplitude is determined by a trade-off between a quick escape from the initial well, a low recrossing probability and a short slide time.
A simple approximation of this optimal potential profile is given by the N-shaped function (denoted B'), only parametrized by the potential barrier height $U_1$, as shown in Fig.~\ref{fig2}~(c). A minimization of~\eqref{tau1Dresc} for this potential profile yields a $\ln n$ scaling of the parameter $U_1$, and the corresponding mean exit time
\begin{equation}
\tau_n^{B'} =\frac{1}{2 \ln n}  + o\left(\frac{1}{\ln n}\right). \label{tau2}
\end{equation}
Although optimal potential B' cannot exhibit an overshoot, it captures the correct scaling of the mean exit time which is, as for constraint A, given by the sliding time. In particular, to leading order in $n$, the mean exit time and the total potential difference $\Delta U = 2 U_1$ still verify $\tau_n^{B'} \Delta U \simeq 1$.
The overshoot structure of optimal potential B can be satisfactorily described by a two-parameter potential ansatz. Once optimized, it turns out that this overshoot structure modifies the subleading order of~\eqref{tau2}, only yielding a slightly smaller mean exit time than with optimal potential B'. 
It is interesting here to note that a profile, reminiscent of our N-shape,
was also reported in a discrete model of molecular transport through nanopores \cite{K11}.


To compare the efficiency of the differently constrained potentials, we use the total potential amplitude $\Delta U = \max_x|U(x)|$ as an index. 
This leads to Fig.~\ref{fig2}~(d). 
According to this criterion, the optimal potential A (Fig.~\ref{fig2}~(a)) is of course the most efficient, but displays discontinuities. By contrast, the optimal potential B (Fig.~\ref{fig2}~(b)) is continuous but only poorly efficient, since it requires a large amplitude due to the over- and undershoot, which nevertheless do not significantly reduce the mean exit time. A good compromise is given by the N-shaped reasonably efficient optimal potential B' (Fig.~\ref{fig2}~(c)), which we realized experimentally as we describe below.

\section*{Experimental design and results}

In order to test whether experimental potentials can be tailored to deliver the predicted speed-ups, we leveraged the ability of a holographic optical tweezer (HOT) to create almost arbitrary intensity and phase-patterns in the focal plane of a microscope~\cite{Curtis2012, Gladrow2019}. In addition, we used a microfluidic device to confine movements of colloidal particles, to a quasi-one-dimensional line, eliminating entropic forces and variations in hydrodynamic friction~\cite{Zwanzig1992, Yang2017}. The motion of colloidal particles is well within the overdamped regime, such that our theory applies. All experiments were carried out by an automated ``drag-and-drop'' routine based on a real-time recognition system, which is able to locate colloidal particles and displace them using individually addressable dynamic holographic traps~\cite{Bowman2014}.

\begin{figure}
 \centering
 \includegraphics[width=87mm]{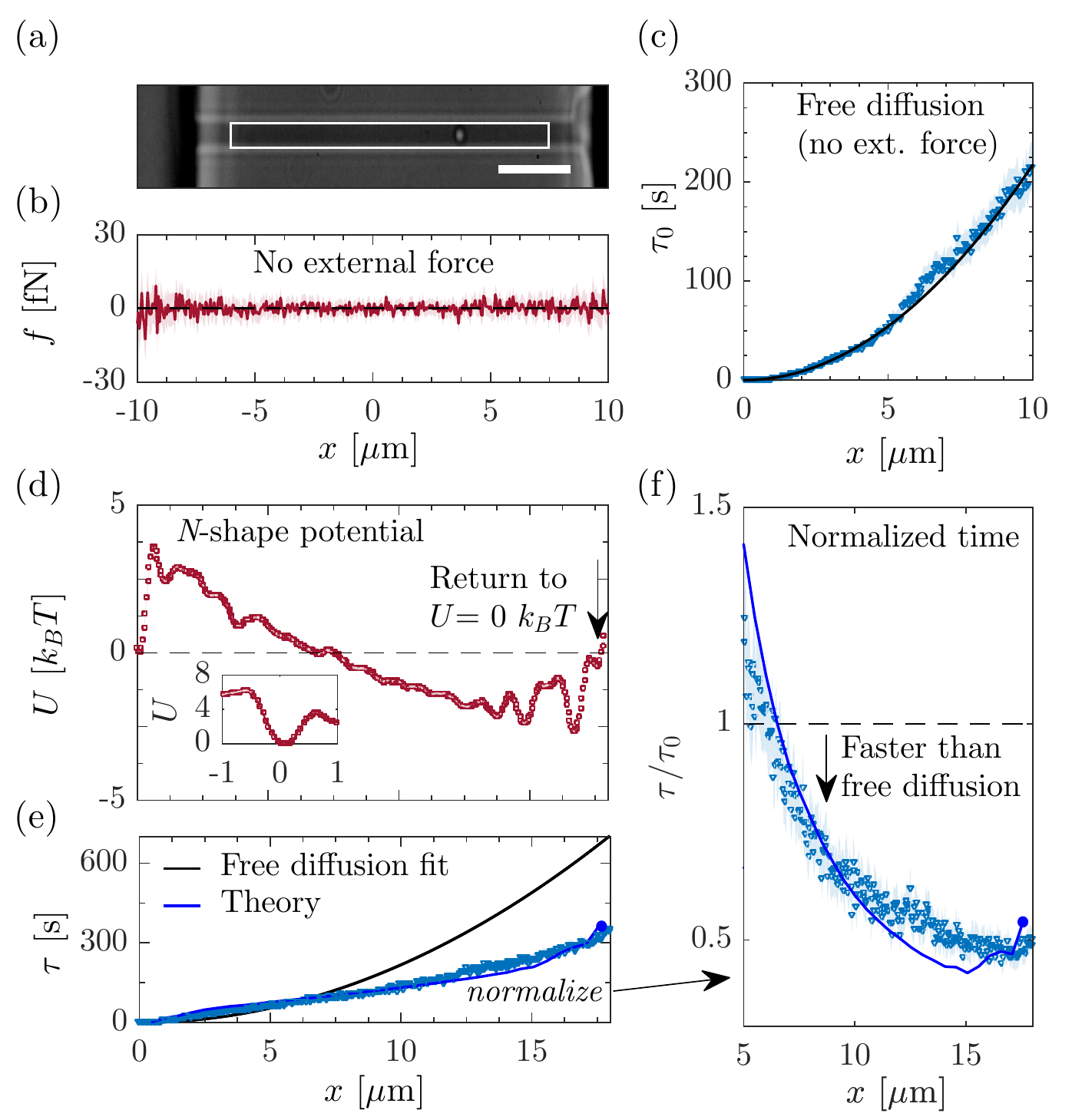}
 \caption{(color online) (a) Picture of microfluidic channel used containing a single particle. The region of interest used subsequently is shown with the frame. (b) Forces along the channel inferred for the zero-potential case. The error-envelope is on the order of the marker size. We here plot the force rather than the potential to highlight the accuracy of our force estimator. (c) First-passage time measured symmetrically from the centre of an interval in the absence of optical forces. The scale bar corresponds to $5 \,\mu$m.  (d) N-shaped potential created by the HOT, corresponding to the rightmost star in Fig.~\ref{fig2}(d) (the other experiment is described in the supplement~\cite{SM}). The inset shows the asymmetric barrier used to approximate the initial reflecting boundary. (e) Measured first-passage times at position $x$ for the same potential compared with the free-diffusion fit. (f) Measured first-passage time at position $x$, normalized by the corresponding free diffusion time $x^2/(2D)$.}
 \label{fig3}
\end{figure}

As a first step, we measured first-passage times $\tau_0$ of a colloid released in the centre of a channel, shown in Fig.~\ref{fig3}(a), without the influence of laser forces (see panel (b)). As the data in Fig.~\ref{fig3}(c) shows, these times adhere closely to theory and scale quadratically with distance. From this data set, we infer a diffusion coefficient of $D=0.23\, \mu\text{m}^2/\text{s}$. 

The holographic parameters necessary to form the right balance of intensity-gradient and phase-gradient forces~\cite{Roichman2008} were found by trial-and-error.
Specifically, the N-shaped potential was created by a combination of a single point trap providing the initial potential well and three line traps with phase-gradients and lengths as specified in the supplement~\cite{SM}. 

The resulting potential landscape $U(x)$ was obtained by integrating forces $f(x)$, which we inferred along the channel from binned displacement statistics $\rho_x(\Delta x) \propto \text{exp}\left[-\frac{(\Delta x -  \Delta t f(x)/(m\gamma))^2}{4 D \Delta t} \right]$. The friction coefficient $\gamma$ was calculated from the measured diffusion coefficient using $\gamma = k_B T /(m D)$. Our passive potential inference works reliably for shallow potential wells $\Delta U< 5\, k_BT$; inference of deeper minima would require intervention~\cite{Juniper2012}. As Fig.~\ref{fig3}(d) shows, the potential largely adheres to the desired N-shape, except for a few wiggles, which are due to optical aberrations and interference (see supplement~\cite{SM}).

The obtained mean first-passage time is plotted in Fig.~\ref{fig3}~(e), as well as the free-diffusion fit. The profile speed-up introduced by the intermediary slide part of this potential is clearly visible. It results in a mean exit time of $336 \pm 19$ s, to be compared to the $684$ s for free diffusion, hence we obtain a speed-up factor of $2$. This experimental measurement is displayed in Fig.~\ref{fig2}~(d) with a star. The experimental landscape has a lower efficiency than the targeted optimized N-shaped potential. This is caused by aberrations and interference between different holographic elements, but in spite of these imperfections, a significant speed-up is achieved.

\section*{The underdamped regime}

\begin{figure}[h]
\centering
\begin{minipage}{0.45\linewidth}
\includegraphics[height=140pt]{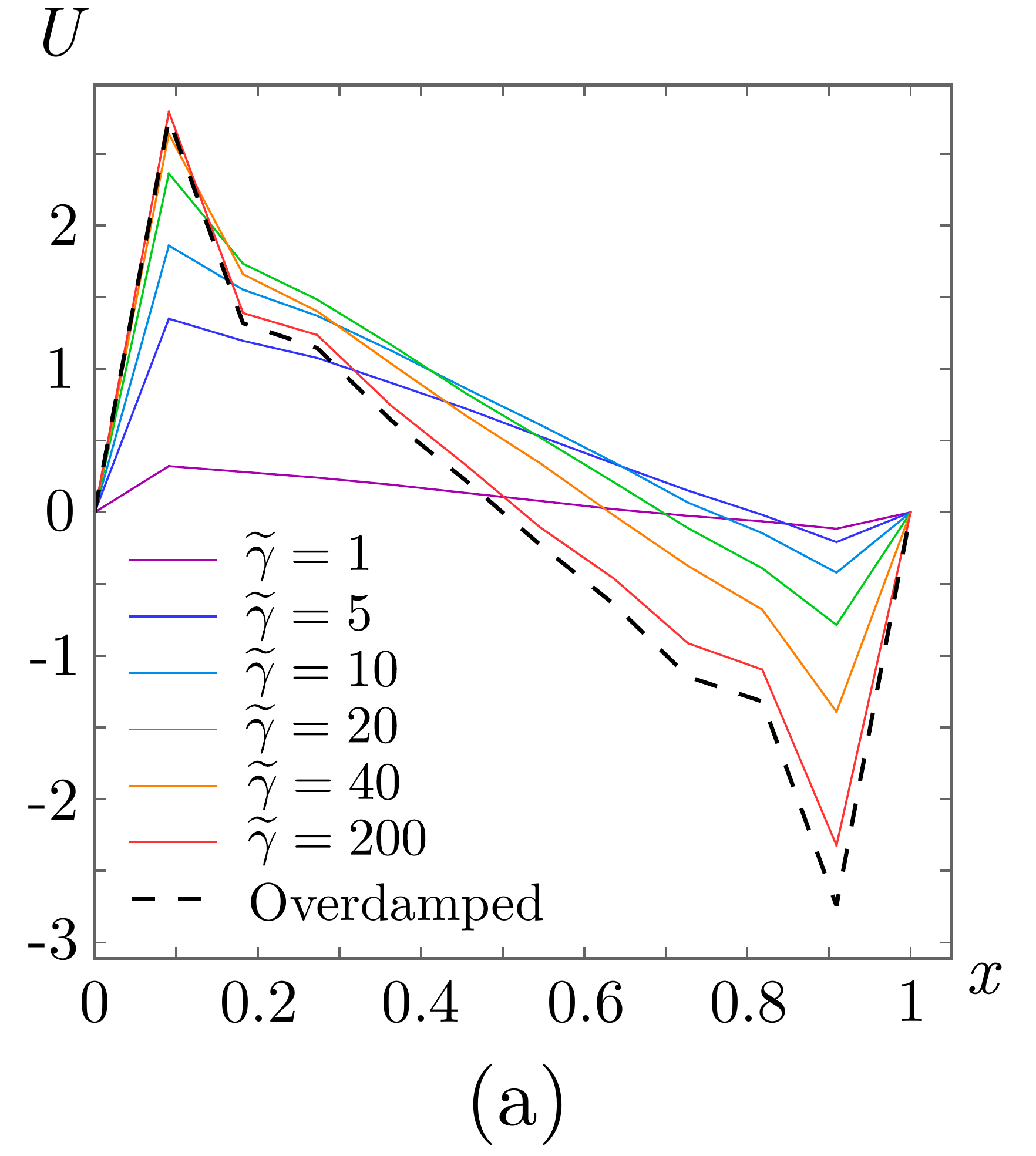}
\end{minipage}
\hfill
\begin{minipage}{0.45\linewidth}
\includegraphics[height=140pt]{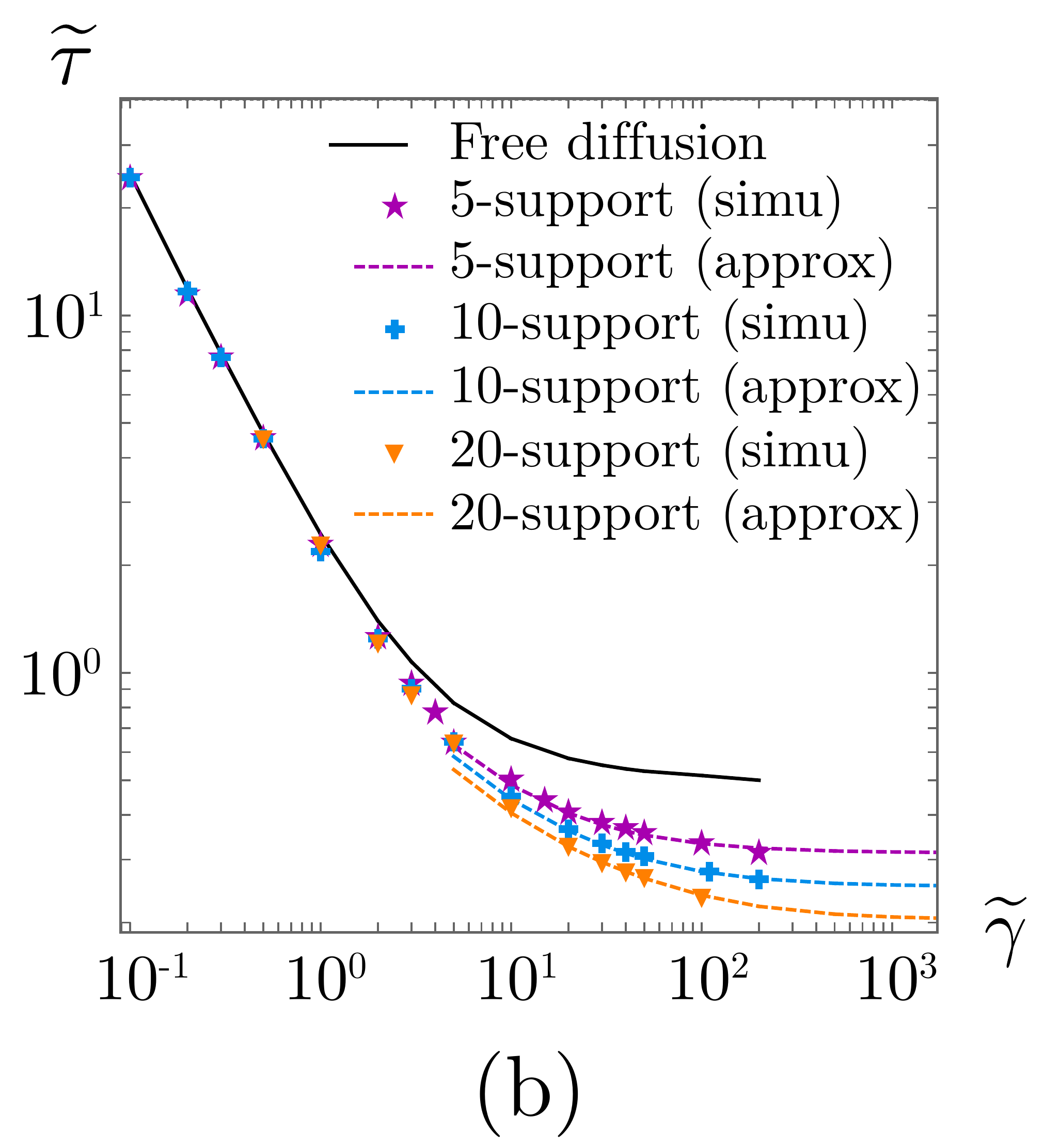}
\end{minipage}
\hfill
\caption{(color online)(a) Optimal 10-support potential profiles for various frictions, compared to the overdamped limit. (b) Evolution of the optimal mean exit time with friction, for 5-, 10- and 20-support potentials, obtained by finite element annealing (see supplement \cite{SM}). Dashed lines correspond to optimization carried out with~\eqref{approxAlexei}.}
\label{under}
\end{figure}

Finally, energy profile optimization raises a conceptual problem. Provided that the constraint is loose enough (large $\Delta U$ or large $n$ for example), the optimal mean exit time becomes sufficiently low, and may leave the range of validity of the overdamped regime \cite{BarratHansen}. Therefore, we need to extend the discussion to the underdamped situation, where the state of the particle is characterized by both its position and velocity (see suppl.~\cite{SM}). Then, inertia matters, and the particle's velocity cannot instantaneously adjust to the force applied. This delay causes a non-trivial response to forcing. To proceed,
we introduce some additional rescalings $\widetilde{v} = v \sqrt{m/(kT)}$ and $\widetilde{\gamma} = \gamma L \sqrt{m/kT}$, where tildes will again be implicit in the following.
Though not consistent with the rescalings on position and velocity, we keep the same rescaling for time as in the overdamped case, for better comparison with this limit. 
Extracting the mean exit time from the statistical description of the underdamped problem is more involved than in the overdamped case, and no general analytical expression is known. Even the free diffusion case requires cumbersome calculations~\cite{Hagan:1989,Masoliver:1995}. However, a development in terms of harmonic oscillator eigenfunctions can be carried out~\cite{Risken}. Keeping the first two orders, we find (see suppl. \cite{SM})
\begin{align}\label{approxAlexei}
&\tau \simeq  \int_0^1 dx \int_x^1 dz \, e^{U(z)-U(x)} 
+ \frac{1}{\gamma}  \sqrt{\frac{\pi}{2}} \int_0^1 dx \, e^{-U(x)} \nonumber \\
&+ \frac{1}{\gamma^2} \left[ - \frac{U'(1)}{2} \int_0^1 \!\! dx \, e^{-U(x)} + \int_0^1 \!\! dx \! \int_x^1 \!\! dz \, U'^2(z) e^{U(z)-U(x)} \right],
\end{align}
here for an initially thermalized particle starting on the reflecting boundary. As expected, the zero-order term corresponds to the overdamped expression~\eqref{tau1Dresc}. Then, the first order term in $1/\gamma$ penalizes negative parts of the potential and favors positive ones, resulting in an antisymmetry breaking of the optimal potential for finite friction. As for the second order term, its first part favors negative increasing profiles near $x=1$, whereas its second part, that is mainly significant around the maximum of the profile, favors height reduction of the first barrier as well as its bending. 

To test this theoretical prediction and work out arbitrary damping, we implement a simulated annealing optimization coupled to a finite element method (see suppl.~\cite{SM}).  
In order to facilitate comparison with the overdamped limit, we restrict the optimization to $n$-support functions. Fig.~\ref{under}(a) shows how a decrease in friction impinges on the optimal potential profile. It confirms the expectation based on~\eqref{approxAlexei}, such as the breakdown of antisymmetry, reduction of the amplitude of the optimal profile, and bending of the profile around its maximum with the disappearance of the overshoot.
We compare the resulting optimized mean exit time with the free case on Fig.~\ref{under}(b). Interestingly, the profile speed-up does extend beyond the overdamped limit. However, its efficiency decreases when friction decreases. In very underdamped situations, despite the momentum gained in the intermediary slide part, the cost for well escape becomes prohibitive. Our results indicate that in the case of vanishing friction, the optimal shape will converge to a constant potential. The profile speed-up is therefore most relevant in the moderately damped to overdamped regimes. 

\section*{Conclusion}
To conclude, we have studied profile speed-up of a Brownian particle by an energy-neutral
potential barrier.
We optimized this process under two complementary constraints, which either bound the potential directly or require regular spatial discretization. From the optimal potentials obtained, we constructed an efficient experiment-friendly profile that we implemented using a combination of optical and microfluidics techniques. 
We were thereby able to accelerate the exit dynamics of a colloid in a narrow channel by a factor two. Moreover, the profile speed-up is not specific to overdamped systems and is observed, though with lower magnitude, at arbitrary damping. Altogether, the profile-induced speed-up then appears to be robust and relevant in a large range of friction values. Finally, although the emphasis was here in the one dimensional setting, our results extend to higher dimensions, as discussed in the supplementary material~\cite{SM}. While a comprehensive analytical treatment of this problem beyond the overdamped limit remains a considerable theoretical challenge, our results anyhow require a rethink of the seemingly settled problem of reaction rates and Brownian transport.

\showmatmethods{} 

\acknow{We thank O. Giraud, Y. Tourigny and F. van Wijland for useful discussions and critical reading of the manuscript.
We acknowledge funding from the Investissement d'Avenir LabEx PALM program (Grant No. ANR-10-LABX-0039-PALM). The research leading to these results has also received 
funding from the European Union’s Horizon 2020 research and innovation program under ETN Grant No. 674979-NANOTRANS. U.F.K. acknowledges funding from an ERC Consolidator Grant (DesignerPores 647144).}

\showacknow{} 

\bibliography{MET}

\newpage

\onecolumn

\begin{center}
{\Large \bf Supplementary material for: ``Optimizing 
Brownian escape rates by potential shaping''}
\end{center}

Starting with setting the framework in section \ref{sec:framework} and being interested in mean-first passage times, we bring to the fore the key feature 
of potential antisymmetry. We then present in  Section \ref{sec:exp} the experimental setup and techniques.
We show in section \ref{sec:anti} that for overdamped dynamics,
the optimal potential is necessarily antisymmetric, as soon as the constraint on the potential is compatible with such a property. Second, we optimize in section \ref{sec:opti}
the potential within constraint A (bounded potential). We then show in section \ref{sec:beyond1D} that the escape time boost is not specific 
to the one dimensional setting on which our attention is mainly focussed, but also holds in higher dimensions.
We finally present the general underdamped framework,
with analytical asymptotic results (section \ref{sec:underanalytics}) and the Finite Element Method technique used 
for numerical resolution (section \ref{undernumerics}).

\section{The general framework and a key property of the mean first passage time}
\label{sec:framework}

The whole analysis is performed at the level of the Langevin equation \cite{Gardiner2009,VanKampen}
for the motion of a Brownian object in the force field stemming from an external static potential $U(x)$. We mostly focus 
on the one dimensional formulation (see Section \ref{sec:beyond1D} otherwise), for which the position $x$ of a `particle' with mass $m$ obeys 
\begin{equation}
m\ddot x \,=\, -\gamma m \dot x \, - U'(x) \, +\, m\gamma \, \sqrt{D} \, \eta(t)
\label{eq:Langevingen}
\end{equation}
where $\gamma$ denotes the friction coefficient, $U(x)$ the potential in which the particle evolves, $D=k_B T/(m\gamma)$ the diffusion coefficient, and $k_B T$ is thermal energy.
The random term $\eta(t)$ is a standard delta-correlated noise: $\langle\eta(t) \eta(t') \rangle  = 2 \delta(t-t')$. Time and space derivatives are denoted by a point and a dash respectively.
In the overdamped limit, met with large friction coefficients, the inertial term becomes negligible,
and the equation of motion simplifies into:
\begin{equation}
\dot{x}=-\frac{1}{m \gamma} U'(x) +\sqrt{D} \, \eta(t),
\end{equation}
where $1/(m\gamma)$ is the mobility.
Statistical properties of the $x$-evolution are enclosed in the probability density function (pdf) $P(x,t)$
to find the particle at the point $x$ at time $t$, starting from a given initial state. The pdf obeys the Smoluchowski equation
\begin{equation}
\partial_t P = \frac{1}{m\gamma} \partial_x(P U') + D \, \partial_x^2 P.
\end{equation}
In the more general underdamped framework associated to Eq. \eqref{eq:Langevingen}, the pdf $K(x,v,t)$ (such that $P=\int K dv$) obeys the so-called
Kramers equation \cite{VanKampen}, given in Eq. \eqref{kramers} below. 

\begin{figure}[h!]
\centering
\includegraphics[width=350 pt]{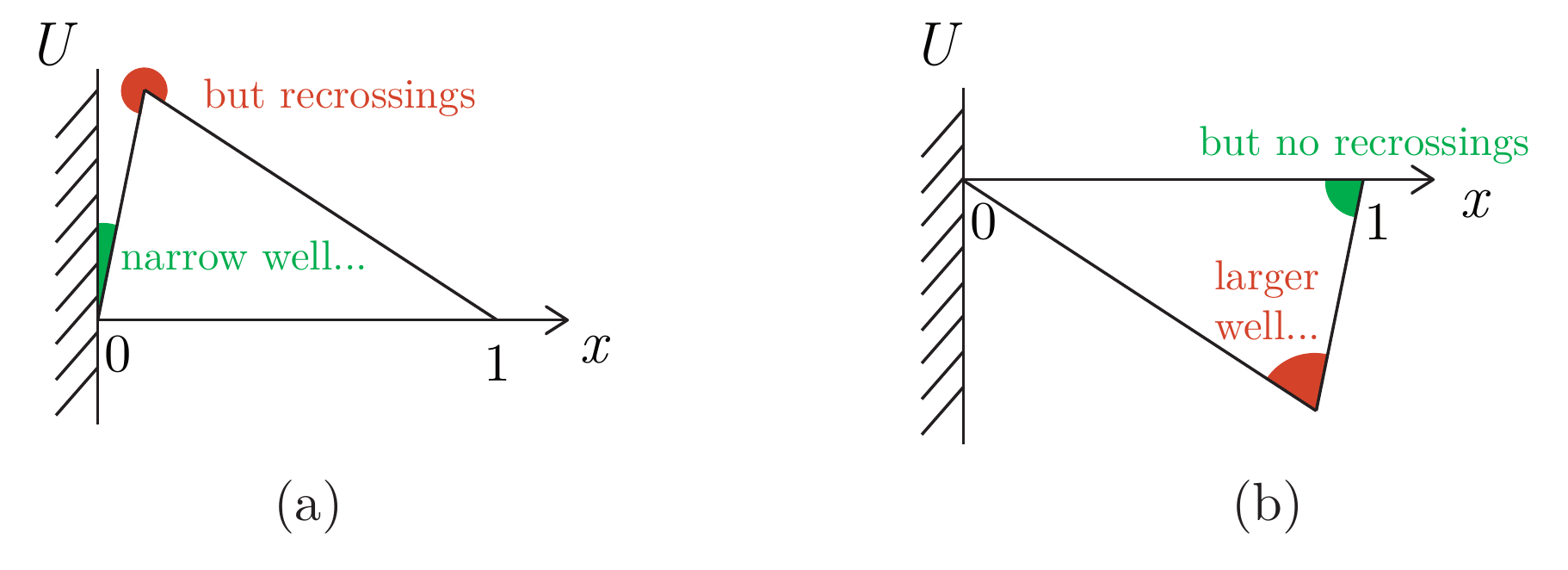}
\caption{Illustration of the generic identity \eqref{eq:antisym}. These two potentials yield the same mean exit time, stemming from a remarkable compensation between the difference of escape time from the two wells and the recrossing occurrence. The particle starts from $x=0$ at the reflecting wall and is absorbed at $x=1$.}
\label{antisym}
\end{figure}

For a given static external potential $U(x)$, a classic argument \cite{VanKampen,BarratHansen} yields the mean first-passage
time $\tau[U(x)]$ provided by Eq. \eqref{tau1Dresc} in the main text. Before embarking in the analysis, it is worth discussing the ensuing interesting
antisymmetry property of the mean first passage time, that plays a fundamental role in our treatment \cite{Palyulin:2012}:
\begin{equation}
\tau[U(x)] \,=\,\tau[-U(1-x)] .
\label{eq:antisym}
\end{equation}
This is a mere consequence of Eq. \eqref{tau1Dresc}; it implies that the two profiles shown in Fig. ~\ref{antisym} lead to the same 
mean first passage time at $x=1$. While this may be surprising at first, it appears that it is completely equivalent to first get out of the well and then slide to the exit 
(Fig.~\ref{antisym}(a)) or first slide down and then climb a steep slope (Fig.~\ref{antisym}(b)). Indeed, if the slide part is identical between (a) and (b), the well part is different. In (a), the particle is blocked leftwards by a reflecting wall, acting like an infinitely high potential, whereas in (b), the well is less confining, because of its left part, which is the slide. Following Kramers-like phenomenology subsumed in Fig. \ref{fig1} in the main text, the particle should (and does) escape quicker from the most confining well of (a). However, once escaped from the well, the particle can fall back down into it in (a) (recrossings) whereas it cannot in (b) as the top of the well is absorbing. 
It turns out that the two effects (well escape and recrossing) exactly compensate and lead to the identity \eqref{eq:antisym}.

\section{Experiments}
\label{sec:exp}

We used an Ytterbium fiber laser (YLM-5-1064-LP, IPG Photonics) at $2.5$ W power with a wavelength of $1064$ nm. To shape the beam, we used a liquid crystal SLM (LCOS X10468, Hamamatsu) with a refresh rate of $60$ Hz. Further details about the setup can be found in previous publications~\cite{Pagliara2013}. The colloidal particles had a nominal diameter of $350$ nm, consisted of polystyrene and are commercially available from Polysciences Inc. The colloids were dispersed in a measurement buffer (3 mM KCl, $0.5$xTris at pH 8) and sonicated prior to the experiment. The microfluidic mask was cast in Polydimethylsiloxane (PDMS) and featured channels with dimensions of $25\times 1 \times 1 \, \mu$m. Both ends of the channel are connected by macroscopic channels to swiftly equilibrate pressure differences.

As stated in the main text, the experimental N-shaped potential represented on 
Fig.~\ref{fig3}(d), corresponding to the rightmost star of Fig.~\ref{fig2}(d), was created by a combination of a single point trap providing the initial potential well and three line traps with phase-gradients and lengths given by the following table.
\begin{figure}[H]
\centering
\begin{tabular}{ l | c | c | c | c  }
 
  Type & $I$ & $x_c$ & $L$ & $p$ \\
  \hline
  Line & $0.2$ & $-0.91\, \mu$m& $2\, \mu$m & $1$ \\
  Point & $0.09$ & $0\, \mu$m & N.A.  &N.A.\\
  Line & $1.1$ & $8.05\, \mu$m & $16\, \mu$m & $0.1$ \\
  Line & $0.6$ & $17.45 \, \mu$m & $3\, \mu$m & $-1$\\
\end{tabular}
\caption{Parameters of the {\em Red tweezers program} \cite{Bowman2014} used to create the N-shaped barrier}
\end{figure}
\noindent $I$ denotes the relative intensities, $x_c$ represents the centre of the respective trap, $L$ stands for the length of the line trap, while $p \in [-1,1]$ is a relative measure for the phase-gradient of a line trap.

\begin{figure}[h!]
\centering
\includegraphics[width=350pt]{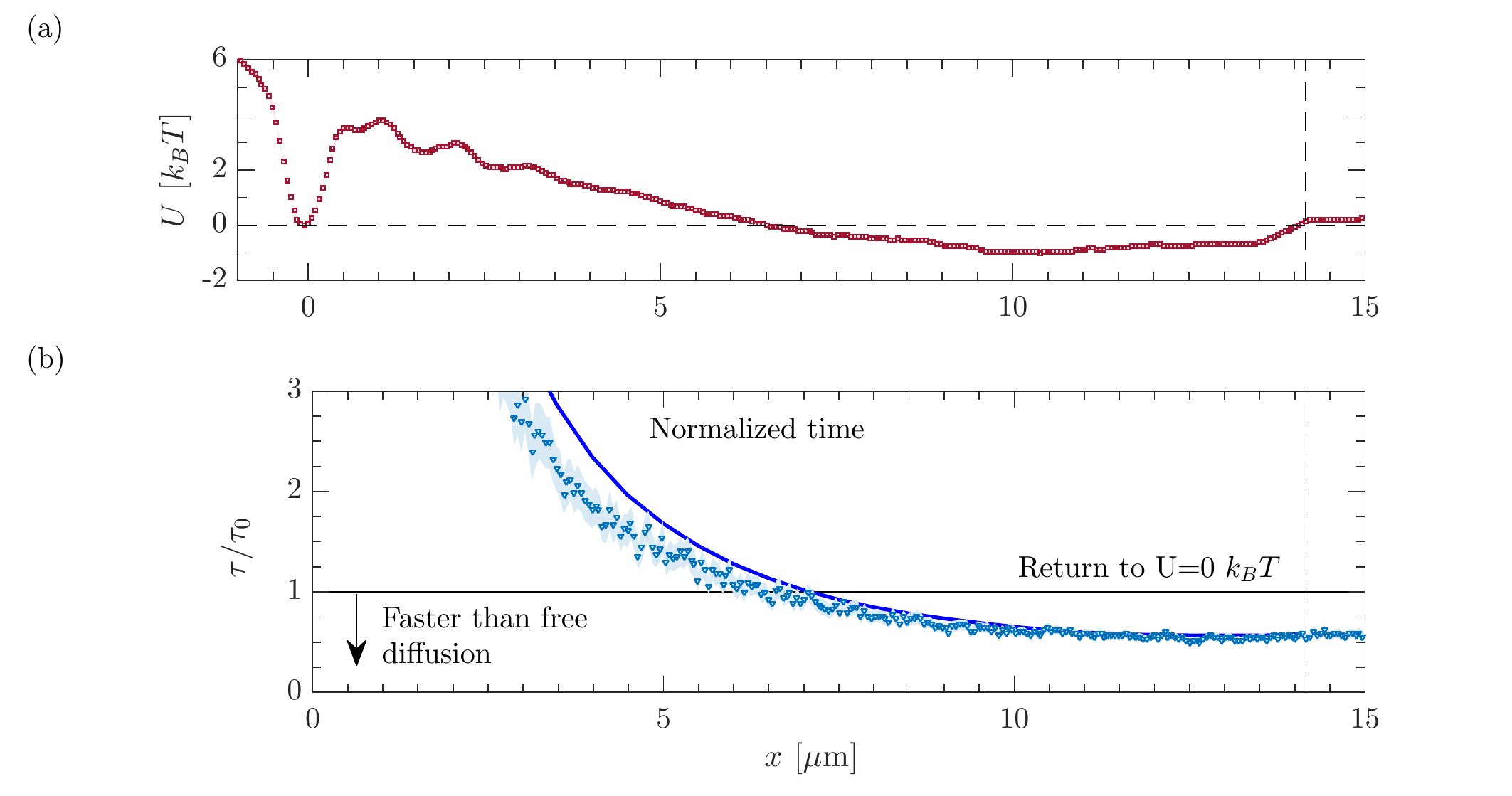}
\caption{(a) Experimental potential profile corresponding to the leftmost star on Fig.~\ref{fig2}(d) of the main text. The interval of interest is between $x=0$ and the vertical dashed line. (b) Measured first-passage times for the potential profile (a) normalized by the zero-potential fit $\langle \tau_0 \rangle$. The continuous curve is the theoretical calculation of $\tau[U]$
following from Eq. (\ref{tau1Dresc}) in the main text, 
where $U$ is the measured potential shown in panel a).
Both (a) and (b) share their $x$ axis.}
\label{figS2}
\end{figure}

This combination of parameters is the one with which we obtained the lowest experimental rescaled mean exit time. We also report another experimental potential that yields a slightly higher mean exit time, but presents a better efficiency (that is to say a better balance acceleration - potential amplitude). This second potential profile, represented in 
Fig.~\ref{figS2}, corresponds to the leftmost star on Fig.~\ref{fig2}(d) in the main text. It is less N-shaped than the experimental profile of Fig.~\ref{fig2}(d), but displays less interference oscillations, preventing the colloid to lose time in intermediary potential wells before exiting.

\section{Antisymmetry of the optimal potential}
\label{sec:anti}

In the overdamped regime, the mean exit time is given by Eq. (\ref{tau1Dresc}) in the main text and 
is the same for a general potential $U(x)$ and for $-U(1-x)$, see section \ref{sec:framework}. Using this property, we can write
\begin{equation}
\tau^2[U(x)]=\tau[U(x)] \; \tau[-U(1-x)]= \int_0^1 dx \, e^{-U(x)} \int_x^1 dy \, e^{U(y)} \times  \int_0^1 dx' \, e^{U(1-x')} \int_{x'}^1 dy' \, e^{-U(1-y')}.
\end{equation}
We use the Cauchy-Schwarz inequality for the following scalar product
\begin{equation}
\langle f|g \rangle = \int_0^1 f(x) g(x)\, dx
\end{equation}
and the associated norm $||f||=\sqrt{\langle f|f \rangle}$. This yields
\begin{equation}
\tau^2[U(x)] \geqslant \left( \int_0^1 dx \, e^{-\frac{U(x)-U(1-x)}{2}} \sqrt{\int_x^1 dy e^{U(y)} \int_x^1 dy' e^{-U(1-y')}} \right)^2.
\end{equation}
We invoke again the Cauchy-Schwarz inequality for the integral scalar product on $[x,1]$ to obtain
\begin{equation}
\tau^2[U(x)] \geqslant  \left( \int_0^1 dx \, e^{-\frac{U(x)-U(1-x)}{2}} \int_x^1 dy \, e^{\frac{U(y)-U(1-y)}{2}} \right)^2 = \tau^2 \left[ \frac{U(x)-U(1-x)}{2} \right].
\end{equation}
This function $[U(x)-U(1-x)]/2=U_a(x)$ is the original potential $U(x)$, anti-symmetrized with respect to $x=1/2$.  It means that
\begin{equation}
\tau^2[U(x)] \geqslant \tau^2[U_a(x)],
\end{equation}
namely that the mean exit time associated with a potential $U$ is always larger than the mean exit time associated with its anti-symmetrized version $U_a$. So, provided that the ensemble of potentials allowed by the regularization constraint is stable under this antisymmetry operation (\textit{i.e.} if $U$ is in this ensemble, so is $U_a$), 
the optimal potential is necessarily antisymmetric itself.

\section{Optimization within constraint A}
\label{sec:opti}

\begin{figure}[h!]
\centering
\includegraphics[width=180pt]{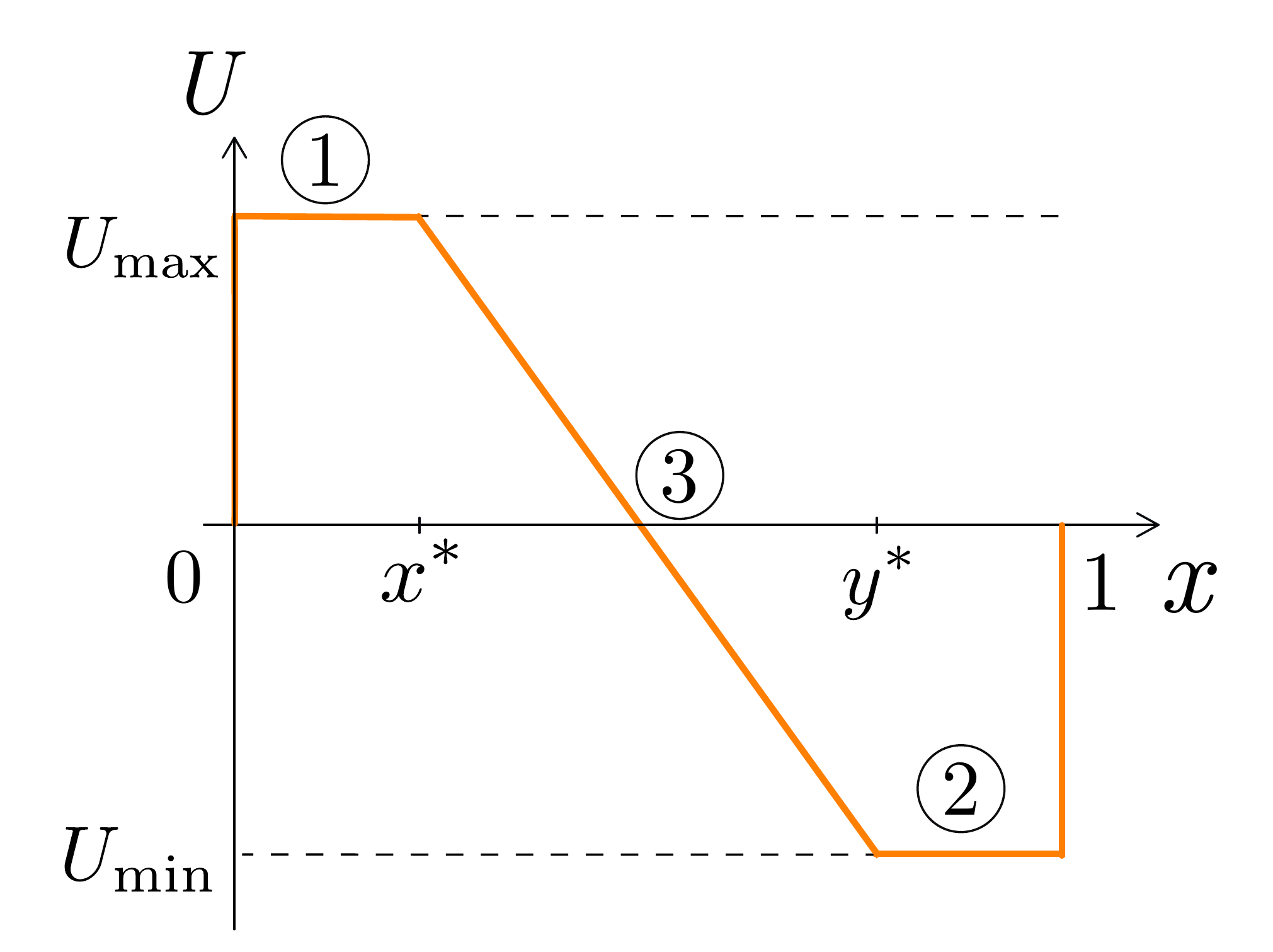}
\caption{Sketch of optimal potential A, with illustration of regions \textcircled{1}, \textcircled{2} and \textcircled{3}.}
\label{figS1}
\end{figure}

With constraint A, the potential is such that for all $x \in [0,1]$, $U_{\rm min} \leqslant U(x) \leqslant U_{\rm max}$. The optimal profile A that we are looking for can be split into three regions (see Fig.~\ref{figS1}): region \textcircled{1} where it is equal to the upper bound $U_{\rm max}$, region \textcircled{2} where it is equal to the lower bound $U_{\rm min}$ and region \textcircled{3} where it is not constrained ($U_{\rm min} < U(y) < U_{\rm max}$). A priori, these three regions can be non-connected. In order to determine the optimal profile A, we use variational calculus, and compute
\begin{equation}
 \delta \tau = \int_0^1 dx \int_0^1 dy \, \Theta(y-x) e^{U(y)-U(x)} (\delta U(y) -\delta U(x))
\end{equation}
where $\Theta$ is the Heaviside step function. Outside of the constrained regions \textcircled{1} and \textcircled{2}, this variation $\delta \tau$ is zero at the optimum, yielding
\begin{equation}\label{Svar1}
e^{U(y)} \int_0^y dx \, e^{-U(x)} = e^{-U(y)} \int_y^1 dx \, e^{U(x)}.
\end{equation}
We differentiate once this equation with respect to $y$ to get
\begin{equation}\label{var2}
\int_0^y dx \, e^{-U(x)}=-\frac{e^{-U(y)}}{U'(y)}
\end{equation}
and once more
\begin{equation}\label{var3}
\frac{d}{dy}\left(\frac{1}{U'(y)}\right)=0.
\end{equation}
Equation~\eqref{var3} indicates that the optimal potential A is linear in region \textcircled{3}, with a negative slope as follows from Eq.~\eqref{var2}. 
It can be shown that optimal potential A displays the sequence region \textcircled{1}, region \textcircled{3} and region \textcircled{2}. Regions \textcircled{1}, \textcircled{2} and \textcircled{3} are therefore connected and the optimal potential has the generic shape represented in Fig.~\ref{figS1}. As for the abscissas $x^*$ and $y^*$ of the intersection between these three regions, they can be extracted straightforwardly from Eq.~\eqref{Svar1} and~\eqref{var2}.

\section{Beyond one spatial dimension}
\label{sec:beyond1D}

Our problem is readily generalized to an arbitrary space dimension $d$. We focus on the mean time that an overdamped walker needs to escape from a (hyper)sphere, starting from its center. 
We consider a potential with rotational symmetry $U(r)$, and $P(r,t)$ the associated probability density function of the particle. The latter obeys a Fokker-Planck equation
\begin{equation}
\partial_t P=\frac{1}{\gamma \, r^{d-1}} \partial_r\left( r^{d-1} P \partial_r U \right) + \frac{D}{r^{d-1}} \partial_r \left( r^{d-1} \partial_r P \right).
\end{equation}
A derivation using standard techniques leads to the general expression in dimension $d$
\begin{equation}\label{taud}
\tau=\int_0^1 dr \, r^{d-1} e^{-U(r)} \int_r^1 \frac{dr'}{r'^{d-1}} \, e^{U(r')}
\end{equation}
where the mean exit time is rescaled by $R^2/D$ with $R$ the radius of the sphere. This expression can be recast into a one-dimensional-like expression
\begin{equation}
\tau=\int_0^1 dr \, e^{-V(r)} \int_r^1 dr' e^{V(r')}
\label{eq:taudimd}
\end{equation}
with the effective potential $V(r)$ defined as
\begin{equation}
V(r)=U(r)-(d-1) \ln r.
\end{equation}
The logarithmic contribution above invalidates the antisymmetry property discussed in section~\ref{sec:framework}
for one-dimensional potentials. Here, the mean exit time associated with the potential $-U(1-r)$ is not the same as its counterpart for $U(r)$. 

Minimizing $\tau$ in \eqref{eq:taudimd} we obtain the following equation
\begin{equation}\label{vard1}
e^{U(r)} \int_0^r dr' \left( \frac{r'}{r} \right)^{d-1} e^{-U(r')} = e^{-U(r)} \int_r^1 dr' \left( \frac{r}{r'} \right)^{d-1} e^{U(r')}
\end{equation}
and its derivative with respect to $r$
\begin{equation}\label{vard2}
\int_0^r dr' r'^{d-1} e^{-U(r')} = \frac{r^{d-1} e^{-U(r)}}{\frac{d-1}{r} - U'(r)}.
\end{equation}
As previously, taking the derivative with respect to $r$ of equation~\eqref{vard2} leads to the following form for the unconstrained part of the optimal potential 
\begin{equation}
U(r) = (d-1) \ln r - Ar+B,
\end{equation}
with $A>0$. The logarithmic component can be viewed as an entropic force that tends to bias the displacement towards the regions of higher accessible configuration space, 
namely the regions of large radii.

We can investigate more thoroughly the shape of the optimal potential in the particular case of dimension 2. The following analysis could be easily transposed to dimensions 3 and higher. The constants $A$ and $B$ can be determined by continuity of the optimal potential at the transition point between the upper bound and the intermediary portion (in $x^*$) on the one hand, and the intermediary portion and the lower bound (in $y^*$) on the other hand. The generic profile is
\begin{equation}\label{shape2D}
U(r)= \begin{dcases}
U_{\rm max} & \textrm{if } r \in [0, x^*] \\
U_{\rm max} + \ln r - \ln x^* - \frac{U_{\rm max} - \ln x^* - U_{\rm min} + \ln y^*}{y^* - x^*} (r - x^*) & \textrm{if } r \in ]x^*,y^*[ \\
U_{\rm min} & \textrm{if } r \in [y^*,1].
\end{dcases}
\end{equation}
Following the same lines as for the one-dimensional case, i.e. injecting this potential profile into equations~\eqref{vard1} and~\eqref{vard2}, we obtain 
\begin{align}
&x^* = - 2 y^* \ln y^* \label{ylny}\\
& \frac{x^*}{2} = \frac{y^* - x^*}{U_{\rm max} - \ln x^* - U_{\rm min} + \ln y^*} \label{pente}.
\end{align}
Combining these two equations and setting again $\Delta U = U_{\rm max} - U_{\rm min}$, we get
\begin{equation}
2 + \Delta U + \ln \left( \frac{1}{2 \ln \left( \frac{1}{y^*} \right)} \right) = \frac{1}{\ln \left( \frac{1}{y^*} \right)}.
\end{equation}
If we define
\begin{equation}\label{defY}
Y^* = \frac{1}{\ln \left( \frac{1}{y^*} \right)},
\end{equation}
we get the simple yet implicit relation
\begin{equation}\label{eqY}
2 + \Delta U - \ln 2 + \ln Y^* = Y^*.
\end{equation}
This equation has two solutions, one for $Y^* < 1$ and one for $Y^* > 1 $, corresponding respectively to $y^*<e^{-1}$ and $y^*>e^{-1}$. Moreover, by definition, $x^* < y^*$, which implies $y^* > e^{-0.5}$ because of equation~\eqref{ylny}. The only relevant solution is then $Y^* > 1$. 

\begin{figure}[h!]
\centering
\begin{minipage}{0.45\textwidth}
\includegraphics[width=200pt]{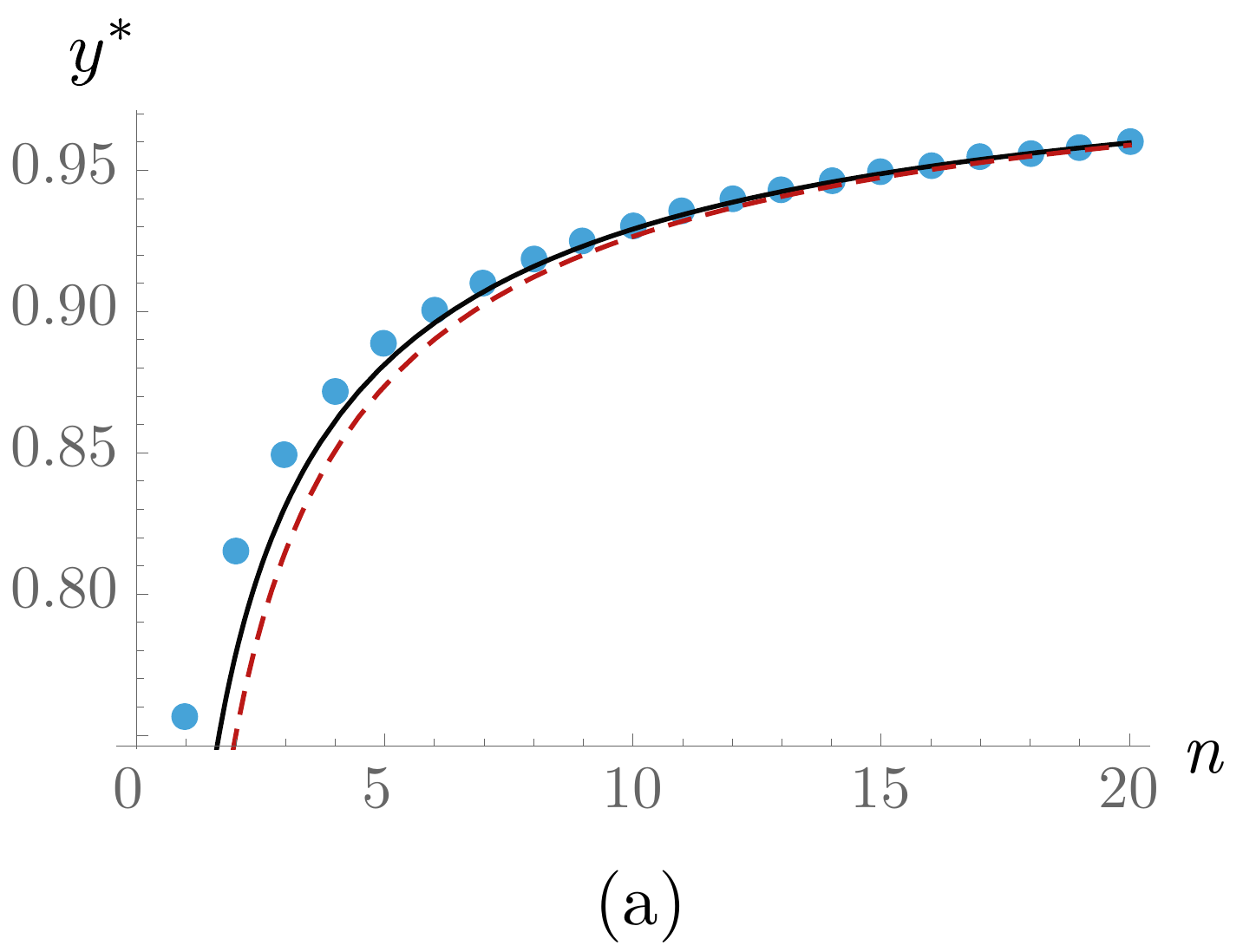}
\end{minipage}
\hfill
\begin{minipage}{0.45\textwidth}
\includegraphics[width=200pt]{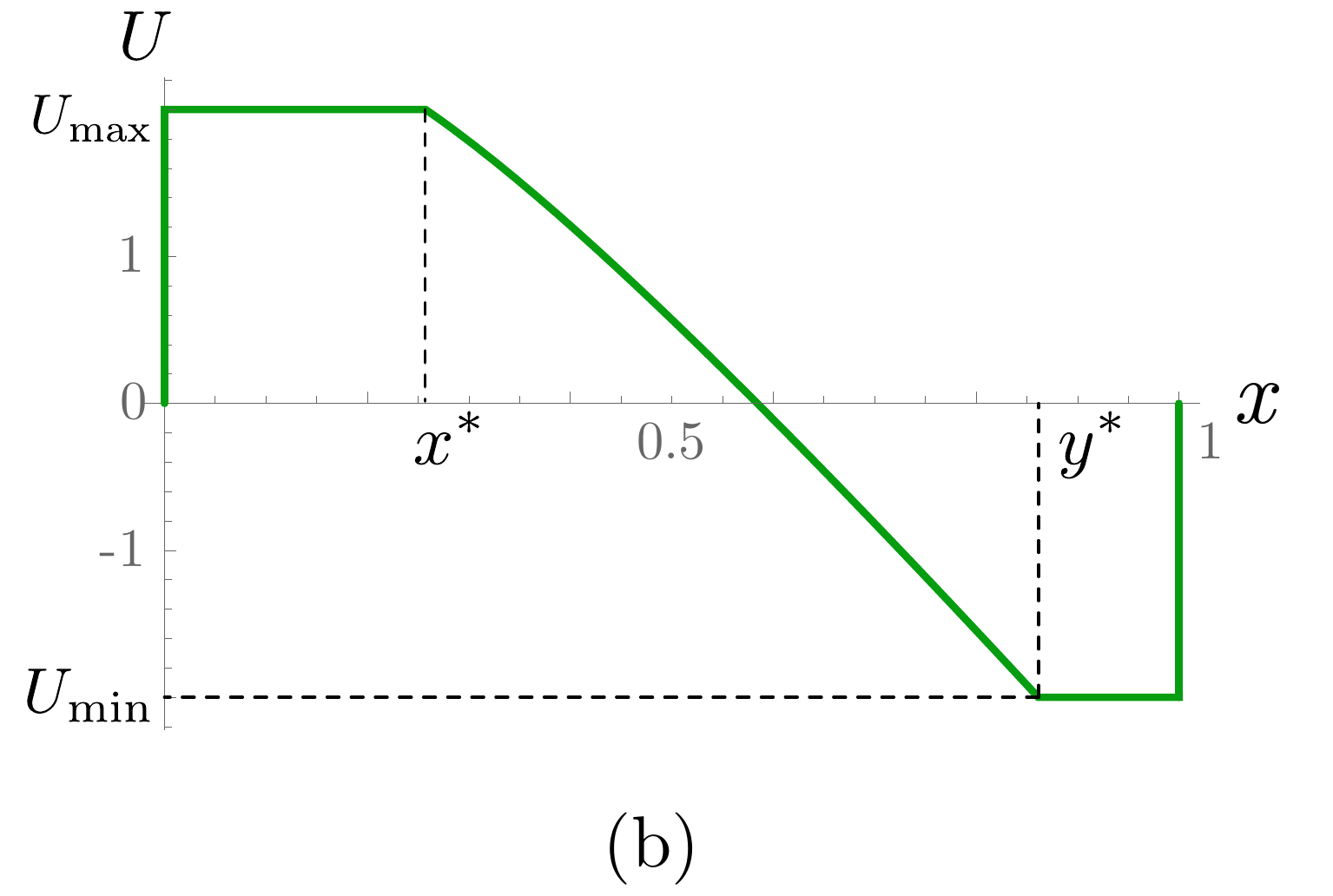}
\end{minipage}
\caption{(a) Comparison between the exact value of $y^*$, obtained by solving numerically Eq.~\eqref{eqY} (blue circles), and the approached expressions~\eqref{ygen} (black line) and~\eqref{ylarge} (red dashed line). (b) Example of optimal potential profile in two dimensions, with the bounds $U_{\rm max}=-U_{\rm min}=2$.}
\label{Uopt}
\end{figure}

In order to solve the implicit equation~\eqref{eqY}, we study its asymptotics when $\Delta \gg 1$. In this case, $Y^*$ is also large compared to 1 and then $\ln Y^* \ll Y^*$. We deduce from equation~\eqref{eqY} the leading term of $Y^*$ and the first corrections
\begin{equation}
Y^* = \Delta U +\ln \Delta U + 2 - \ln 2 + o(1),
\end{equation}
therefore
\begin{align}
& y^* = e^{-\frac{1}{Y^*}}  \simeq e^{-\frac{1}{\Delta U + \ln \Delta U+ 2 - \ln 2}} \label{ygen}\\
& x^* = \frac{2 e^{-\frac{1}{\Delta U + \ln \Delta U + 2 - \ln 2}}}{\Delta U + \ln \Delta U + 2 - \ln 2}.
\end{align}
At large $\Delta U$, this reduces to
\begin{align}
& y^* = 1 - \frac{1}{\Delta U + \ln \Delta U + 2 - \ln 2} \label{ylarge}\\
& x^* = \frac{2}{\Delta U + \ln \Delta U + 2 - \ln 2}.\label{xlarge}
\end{align}
Note that the relation $y^*=1-x^*$ that was verified in one dimension does not hold in two dimensions, even asymptotically. The accuracy of the developments~\eqref{ygen} and~\eqref{ylarge} is tested on figure~\ref{Uopt}(a) as a function of $\Delta U$. The shape of the optimal potential in two dimensions is represented on figure~\ref{Uopt}(b). The asymmetry introduced by the entropic force, that pushes the particle away from the center, has two manifestations: the non-linearity of the intermediary portion and the size of the plateau on $U_{\rm min}$ that is shorter than the size of the plateau on $U_{\rm max}$, see Fig. \ref{Uopt}.

The mean exit time corresponding to this optimal potential can be expressed using the generic shape given in Eq.~\eqref{shape2D} as well as Eqs.~\eqref{ylny} and~\eqref{pente}
\begin{equation}
\tau = \frac{x^* y^*}{4} + \frac{1-y^{*2}}{4}.
\end{equation}
The asymptotics of this expression for potential difference $\Delta U \gg 1$ can be extracted using Eqs.~\eqref{ylarge} and~\eqref{xlarge}
\begin{equation}
\tau = \frac{1}{\Delta U} + o\left( \frac{1}{\Delta U} \right),
\end{equation}
which means that the Heisenberg-like relation $\tau \Delta U  \sim 1$ still holds at large $\Delta U$ in dimensions larger than 1.

\section{Underdamped treatment: analytics}
\label{sec:underanalytics}

Within the underdamped treatment (Eq. \eqref{eq:Langevingen}), a natural rescaling for the velocity is $\widetilde{v} = v \sqrt{m/(kT)}$. Whereas there was only one timescale in the overdamped problem (the diffusion time $t_d=L^2/D$), there is now a second (ballistic) one $t_b=L \sqrt{m/(kT)}$, related to the velocity scale. 
We accordingly define two dimensionless times, 
\begin{equation}
\widetilde{t}=\frac{t}{t_d}, \qquad \widetilde{\widetilde{t}}=\frac{t}{t_b}= \widetilde{\gamma} \widetilde{t}.
\end{equation}
Rescaling with $t_b$ is natural for calculations, as it is consistent with the rescalings of space and velocity already chosen. We use the resulting  $\widetilde{\widetilde{t}}$
in the following calculations. However, rescaling with $t_d$ is adapted for comparison with the overdamped case. In particular, within this second rescaling, 
the limit of the mean exit time at large $\gamma$ is a constant ($\widetilde \tau \to 1/2$), making the large damping asymptotics easier to visualize. We adopted this second rescaling in the main text. As previously, we only work with rescaled variables and then drop tildes, except for the time, where we keep the tilde (or double tilde) to avoid confusion. 

The underdamped probability density function $K\left(x,v,\widetilde{\widetilde{t}}\right)$, 
obeys the Kramers equation
\begin{equation}\label{kramers}
 \partial_{\widetilde{\widetilde{t}}} \, K + v \partial_x K - \partial_x U \partial_v K = \gamma \partial_v \left( v K + \partial_v K \right).
\end{equation}
The starting point $x=0$ and the initial velocity $v$, distributed as the equilibrium Gaussian ${\cal N}(v)$ 
\begin{equation}
{\cal N}(v)=\frac{1}{\sqrt{2\pi}} e^{-\frac{v^2}{2}},
\end{equation}
are implicit in the following. Solving this equation for the mean exit time is much more complicated in the underdamped regime and no analytical general result is known. However, the mean exit time can be related to the particle density function. In order to show that, we first introduce the Green function
\begin{equation}\label{defGreen}
G(x,v) = \int_0^{+\infty} d\widetilde{\widetilde{t}} \, K\left(x,v,\widetilde{\widetilde{t}}\right).
\end{equation}
Integrating Kramers equation~\eqref{kramers} over time, we get
\begin{equation}
v \partial_x G - \partial_x U \partial_v G - \gamma \partial_v ( v G + \partial_v G) =  \delta(x) \mathcal{N}(v).
\end{equation}
Then, if we call $F\left( \, \widetilde{\widetilde{t}} \,\right)$ the exit time density, the mean exit time is simply
\begin{equation}
\widetilde{\widetilde{\tau}}=\int_0^{+\infty} d\widetilde{\widetilde{t}} \; \widetilde{\widetilde{t}} \; F\left( \, \widetilde{\widetilde{t}} \, \right).
\end{equation}
This first passage time density is related to the survival probability $S\left(\widetilde{\widetilde{t}}\right)$ through
\begin{equation}
F\left( \, \widetilde{\widetilde{t}}\, \right)=-\frac{dS}{d\widetilde{\widetilde{t}}},
\end{equation}
and the survival probability can be deduced from the pdf
\begin{equation}
S\left( \, \widetilde{\widetilde{t}} \, \right)=\int_0^1 dx \int_{-\infty}^{+\infty} dv \, K\left(x,v,\widetilde{\widetilde{t}}\right).
\end{equation}
Using the definition of Green function and integrating by parts, we arrive at
\begin{equation}\label{tauu}
\widetilde{\widetilde{\tau}} =  \int_0^1 dx \int_{-\infty}^{+\infty} dv \, G(x,v).
\end{equation}

Following Ref. \cite{Risken}, we can decompose the velocity dependence of the Green function on the basis of rescaled harmonic oscillator eigenfunctions $\psi_n(v)$ as follows
\begin{equation}
G(x,v)= \psi_0(v)  \sum_{n} c_{n}(x) \psi_n(v).
\end{equation}
The eigenfunctions $\psi_n(v)$ are related to the usual harmonic oscillator wavefunctions $\psi^H_n(v)$ by 
\begin{align}
\psi_n(v) = 2^{-1/4} \psi_n^H(2^{-1/2} v).
\end{align}
For example, for $n = 0$, we have $\psi_0^H(v) = \pi^{-1/4} \exp\left( -\frac{v^2}{2} \right)$ and 
\begin{align}
\psi_0(v)  = (2 \pi)^{-1/4} \exp\left( -\frac{v^2}{4} \right).
\end{align}
This function is related to the equilibrium velocity distribution ${\cal N}(v)$ through $\psi_0^2(v)={\cal N}(v)$. The mean escape time can be written in terms of $c_0(x)$ as
\begin{align}
\widetilde{\widetilde{\tau}} = \int dx \, dv \, G(x,v) = \int dx  \, c_0(x)
\end{align}
due to orthogonality relations between the $\psi_n$.

The advantage of this representation is that the coupling between neighboring modes $c_n(x)$ is proportional to $\gamma^{-1}$ which allows, for periodic boundary conditions, to write an exact asymptotic expansion in the limit of strong damping $\gamma$.  For our case, these equations read: 
\begin{align}\label{EqH}
\left\{
\begin{array}{c}
 \partial_x c_{1} = \delta(x) \\
 n \gamma c_{n} + \sqrt{n} (\partial_x + \partial_x U) c_{n-1} + \sqrt{n+1} \,  \partial_x c_{n+1} = 0  \qquad \text{for } n\ge 1 .
\end{array}
\right.
\end{align}
In \cite{Risken} these equations are then solved by continued fraction expansion which converges quickly even for small $\gamma$. Unfortunately this approach cannot be used here due to the boundary conditions. While the reflecting boundary conditions can still be written simply in the oscillator eigenfunction basis $c_{n}(0) = 0$ for odd $n$, the absorbing boundary conditions introduce a coupling between all the modes $n$.
To see this, we rewrite the absorbing boundary condition $G(1,v) = 0$ for $v < 0$ in the $\psi_n$ basis
\begin{align}
c_{2n,0}(1) = \sum_{m} S_{nm} c_{2m+1,0}(1)
\end{align}
where the coefficients $S_{nm} $ are given by: 
\begin{align}
S_{nm} = 2 \int_0^{\infty} \psi_{2n}(v) \psi_{2m+1}(v) dv.
\end{align}
The (infinite) matrix ${\hat S} = (S_{nm})$ is orthogonal ${\hat S} {\hat S}^t = {\hat I}$. 

Using properties of Hermite functions \cite{Dvorak:1973}, we found the asymptotic behavior of the first coefficients $S_{0m},S_{1m}$ 
\begin{align}
&S_{0m} \sim \frac{(-1)^{m+1}}{\sqrt{2} \pi^{3/4} m^{5/4}} \\
&S_{1m} \sim \frac{(-1)^m \sqrt{3}}{2 \pi^{3/4} m^{5/4}}.
\end{align}
They decay quite slowly with $m$, thus in principle all the $c_n$ modes are coupled and the property that $c_n$ is of order $\gamma^{-n}$ breaks down. 
However, we found that keeping only the first modes still provides an accurate numerical approximation (even if it is no longer an exact asymptotic development). Retaining the lowest modes, we truncate the hierarchy \eqref{EqH} to 
\begin{align}
\left\{
\begin{array}{c}
\partial_x c_1= \delta(x ) \\
\gamma c_1 + \partial_x c_0 + \partial_x U c_0+ \sqrt{2} \partial_x c_2 = 0 \\
2 \gamma c_2 + \sqrt{2}  \partial_x c_1 + \partial_x U \sqrt{2} c_1 = 0
\end{array}
\right.
\end{align}
and the boundary conditions to: 
\begin{align}
\left\{
\begin{array}{c}
c_1(0) = 0 \\
c_1(1) = S_{00} c_0(1) + S_{10} c_2(1) = \sqrt{\frac{2}{\pi}} c_0(1) + \frac{1}{\sqrt{\pi}} c_2(1).
\end{array}
\right.
\end{align}
Solving this set of linear equations, and using $U(0)=U(1)=0$, we obtain
\begin{align}
c_0(x) = \gamma e^{-U(x)} \int_x^1 dy e^{U(y)} +  \sqrt{\frac{\pi}{2}} \left( 1 + \frac{U'(1)}{\sqrt{2 \pi} \gamma}  \right) e^{-U(x)} - \gamma^{-1} e^{-U(x)} \int_x^1 dy U''(y) e^{U(y)}.
\end{align}
Coming back to the rescaling with $t_d$ used in the main text and in the figures, we get 
\begin{align}
\widetilde{\tau} =  \int_0^{1} dx \int_x^1 dy e^{U(y)-U(x)} + \sqrt{\frac{\pi}{2}} \left( \frac{1}{\gamma} + \frac{U'(1)}{\sqrt{2 \pi} \gamma^2}  \right) \int_0^{1} dx e^{-U(x)} - \frac{1}{\gamma^2} \int_0^{1} dx \int_x^1 dy\; U''(y) e^{U(y)-U(x)}.
\end{align}

\section{Underdamped treatment: numerics}
\label{undernumerics}

\begin{figure}[h!]
\centering
\includegraphics[width=150pt]{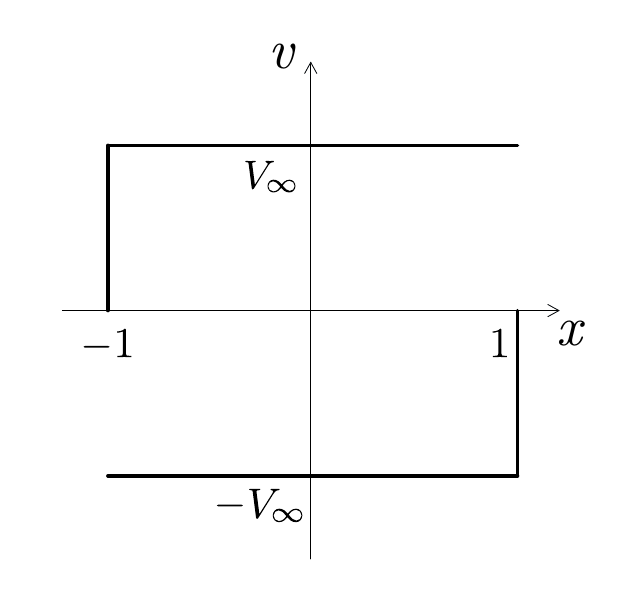}
\caption{Illustration of the domain $\delta \Omega$, constituted of the two portions of thick black line, on which the Green function is 0. The shape of the domain follows from 
Eq.~\eqref{KramGreen}, which contains a second-order derivative in $v$ but only first-order derivatives in $x$. Therefore two boundary conditions are needed in $v$, whereas only one is needed in $x$. In $v$, we simply use a cut-off $\pm V_{\infty}$. In $x$, the absorbing point at $x=1$ prevents the particle to have a negative velocity at this point (as it would come from beyond this absorbing point), as well as the absorbing point at $x=-1$ prevents the particle from having positive velocity there.}
\label{domaine}
\end{figure}

The large-$\gamma$ asymptotics derived above only provides a partial picture of the underdamped regime. To explore a larger range of frictions, we resort to numerics to evaluate the mean exit time associated to a general potential profile. In particular, we use finite element techniques to compute the Green function, defined in Eq.~\eqref{defGreen}, and that obeys
\begin{equation}\label{KramGreen}
v \partial_x G - \partial_x U \partial_v G - \gamma \partial_v ( v G + \partial_v G) = \delta(x) {\cal N}(v).
\end{equation}
To take the reflecting boundary conditions at $x = 0$ into account, we symmetrize the domain explored by the Brownian particles and add an absorbing wall at $x = -1$. The boundary conditions for $G$ can then be written as $G(\delta \Omega) = 0$ where $\delta \Omega$ is the domain shown on Fig~\ref{domaine}. 
The $G(x, v \rightarrow \pm \infty) = 0$ condition is replaced by a hard wall cut-off at $G(x, v=\pm V_\infty) = 0$.
The peculiar shape of this domain is due to the fact that Kramers equation contains a derivative of order two in $v$ but only first order derivatives in $x$~\cite{Araujo:2014}.

To find $G$ using a finite element method, we adopt the weak formulation given below where $\phi(x, v)$ is a trial function, and integration spreads over the entire domain in Fig.~\ref{domaine} 
\begin{equation}
\int \left\{\phi \left( v \partial_x G \,-\, \partial_x U \, \partial_v G \right)  \,+\, \gamma v \,G\, \partial_v \phi  \,+\, \gamma \, \partial_v \phi\, \partial_v G \right\} dv \,=\, 
\delta(x)\, \int \phi  \, \mathcal{N} dv,
\end{equation}
as follows from Eq. \eqref{KramGreen}.
We implemented this weak form using the FreeFem++ simulation package~\cite{Hecht:2012}. To test the numerical accuracy of the solutions obtained, 
we checked the conservation of current. For the exact solution, the current $J(x) = \int v G(x,v) dv$ is independent of $x$ for all $x$ on the same side of the source point. For our simulation parameters, the current was typically conserved with an  accuracy $\sim 5\%$, with a maximal error of $15\%$ for the lowest $\gamma$ with which the convergence is the most difficult
to achieve.

Once the mean exit time estimated through this finite element procedure for a given $n$-support potential profile, we use a simulated annealing algorithm, whose principle is the following. 
We pick at random one of the $n$ vertices of the potential and change its $U$ component by a random quantity in a predetermined range. This change is then accepted or rejected with a Metropolis procedure: if the new mean exit time, computed with finite elements, is smaller than the previous one, then the change is accepted. If it is higher, it is accepted with a probability 
\begin{equation}
P_{\rm acc} \left(\tau_{\rm new},\tau_{\rm old},j \right) = \exp\left(-\frac{\tau_{\rm new}-\tau_{\rm old}}{T_j}\right)
\end{equation}
where $T_j$ acts as a (computational) temperature, and is decreased as $1/j^2$ as the iteration step $j$ increases. The algorithm is then stopped after a predetermined sufficiently high number of unsuccessful iterations (during which the mean exit time does not decrease).

\end{document}